\documentclass[aps, pra, 10pt,  amssymb, amsmath, superscriptaddress, tightenlines, twocolumn, notitlepage, longbibliography,citeautoscript] {revtex4-2}
\usepackage{graphicx}
\usepackage{hyperref}
\hypersetup{
    bookmarks=false,         
    unicode=false,          
    pdftoolbar=false,        
    pdfmenubar=true,        
    pdffitwindow=false,     
    pdfstartview={FitH},    
    pdftitle={},    
    pdfauthor={},     
    pdfsubject={},   
    pdfcreator={},   
    pdfproducer={}, 
    pdfkeywords={}, 
    pdfnewwindow=true,      
    colorlinks=true,       
    linkcolor=blue,          
    citecolor=blue,        
    filecolor=magenta,      
    urlcolor=blue           
}
\usepackage{bm}
\usepackage{physics}
\usepackage{amsmath}
\usepackage[dvipsnames]{xcolor}

\usepackage{lipsum}

\newcommand{\be}{\begin{equation}}
\newcommand{\ee}{\end{equation}}
\newcommand{\bea}{\begin{eqnarray}}
\newcommand{\eea}{\end{eqnarray}}

\newcommand{\p}{\partial}

\renewcommand{\Re}{{\rm \, Re\,}}
\renewcommand{\Im}{{\rm \, Im\,}}

\renewcommand{\subsection}[1]{\noindent{\bf #1}\hspace{.2in}}

\newcommand{\vstag}{{\bm u}}

\begin{document}
\title{
Nonlinear orbital response across topological phase transition in centrosymmetric materials
}
\author{Margarita Davydova}
\affiliation{Department of Physics, Massachusetts Institute of Technology, Cambridge, MA 02139, USA}
\affiliation{IST Austria, Am Campus 1, 3400 Klosterneuburg, Austria}
\author{Maksym Serbyn}
\affiliation{IST Austria, Am Campus 1, 3400 Klosterneuburg, Austria}
\author{Hiroaki Ishizuka}
\affiliation{Department of Physics, Massachusetts Institute of Technology, Cambridge, MA 02139, USA}
\affiliation{Department of Physics, Tokyo Institute of Technology, Meguro-ku, Tokyo, 152-8551, Japan}

\begin{abstract}
    Nonlinear optical responses are often used as a probe for studying electronic properties of materials. For topological materials, studies so far focused on the photogalvanic electric current, which requires breaking inversion symmetry. In this work we present a theory of orbital current response in inversion-symmetric topological insulators. We find  a symmetry-allowed orbital current response that occurs in centrosymmetric materials under illumination by linearly polarized light. The sign of the dc nonlinear conductivity reflects the $\mathbb{Z}_2$ index and the conductivity \emph{changes sign at the  transition} between trivial and topological insulator phases. We derive an expression for the nonlinear orbital photocurrent for a general class of models with two doubly degenerate bands, and discuss its specific applications in the cases of the Bernevig-Hughes-Zhang model and the 1T' phase of transition metal dichalcogenides. Experimental setups for observation of the orbital current are also discussed.
\end{abstract}

\maketitle

\section*{Introduction}

The topology of band structure manifests itself in a rich variety of transport phenomena. A key quantity in these phenomena is the Berry curvature of electronic bands~\cite{Xiao2010}. The Berry curvature modifies the equations of motion in the low energy limit and is responsible for the emergence of the anomalous velocity which is perpendicular to the external electric field~\cite{Karplus1954}. The effect of modified electron dynamics manifests as intrinsic~\cite{Karplus1954,Nagaosa2010} and extrinsic~\cite{Adams1959,Ishizuka2017} anomalous Hall effects. Recent studies determined the effect of the Berry curvature on low-frequency nonlinear responses, such as the nonlinear Hall effect~\cite{Sodemann2015}, photovoltaic effects~\cite{sipe1,Ishizuka2016}, and magnetoresistance~\cite{Son2013,Kim2014,Ishizuka2019c}. For  degenerate bands, the Berry curvature becomes non-Abelian~\cite{PhysRevLett.52.2111,demler1999non} and also affects the dynamics of Bloch wave packets \cite{shindou2005noncommutative}. It is responsible for the emergence of a dissipationless intrinsic spin Hall current in p-doped semiconductors~\cite{Murakami2004,Murakami2003}.

Optical responses involving interband electron transitions also reflect the topological nature of band structure. In photovoltaics, the shift current~\cite{Kraut1979,Belinicher1980,vBaltz1981,Cook2017,Tokura2018,PhysRevLett.124.087402} has experienced a revival of interest in recent years. Recent studies in Weyl semimetals found large second harmonic~\cite{Wu2017} and photovoltaic~\cite{Ma2017,Osterhoudt2019} responses; first-principle calculations demonstrated that they are related to the Weyl nodes~\cite{Zhang2018}. On the other hand, the quantization of another kind of photovoltaic response, the injection current, has been observed both theoretically~\cite{Juan2017} and  experimentally~\cite{Rees2020}. All these phenomena, which take place in noncentrosymmetric materials, were shown to be related to the Berry curvature of electronic bands.

In contrast, the role of topology in light-matter interactions in \emph{centrosymmetric} materials remains largely unexplored. In such systems, which possess both time-reversal and inversion symmetries, photovoltaic and second-harmonic responses of electric current are symmetry prohibited. In this work, we construct an \emph{orbital current} whose response reflects the topological nature of inversion-symmetric topological insulators~\cite{RevModPhys.82.3045,RevModPhys.83.1057}; in particular, the sign of the orbital current depends on the topological index of  2D topological insulators.  

\begin{figure}[thp] 
	\includegraphics[width= \columnwidth]{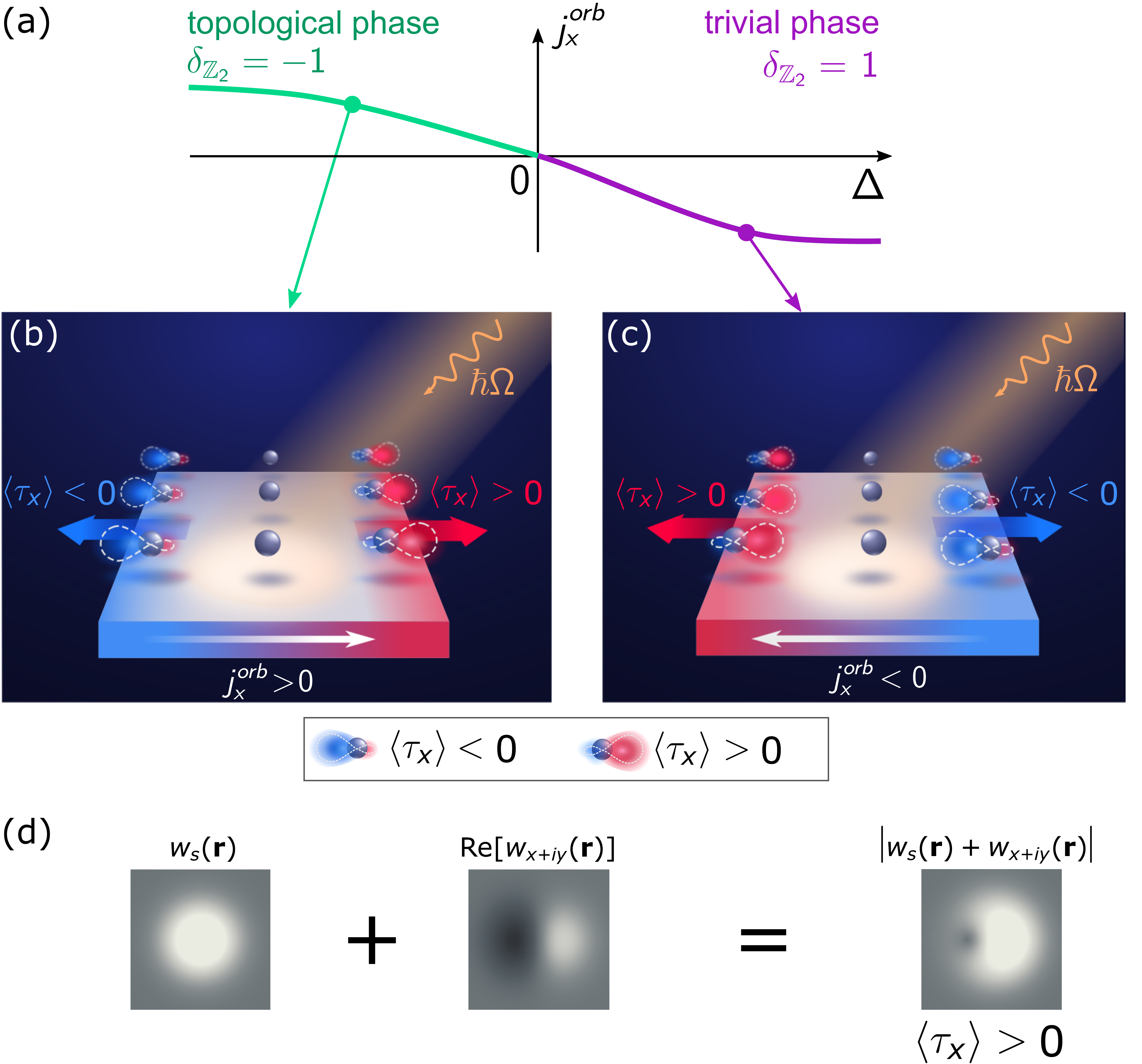} 	
	\caption{%
	(a)  The orbital current  changes its sign as the  mass term (gap) reverses sign at the topological phase transition.  Schematics of the $\tau_x$-orbital separation as a result of the flow of orbital current due to irradiation are illustrated below (b)  and  above (c) the topological phase transition. (d) Construction of an orbital with a fixed eigenvalue of $\tau_x$ operator requires an addition of $w_s$ and $w_{x\pm i y} \equiv w_{p_x\pm i p_y}$ Wannier orbital states. This results in displacement of the electron density  in $x$-direction with respect to the atomic center.} \label{fig1}
	\end{figure}

To explicate the nature of this photo-induced orbital current, we present a general theory for semiconductors with two valence and two conduction bands. Focusing on the case of linearly polarized light, we find at the topological phase transition, the nonlinear conductivity corresponding to the orbital current changes sign, which reflects the change in the $\mathbb{Z}_2$ topological index [Fig.~\ref{fig1}a].
For the Bernevig-Hughes-Zhang (BHZ) model~\cite{Bernevig1757}, an effective model for HgTe/CdTe quantum wells, the orbital current corresponds to a transport of intracell electric polarization. Thus, accumulation of the orbital current will produce ferroelectric polarization at the two ends of the material [Fig.~\ref{fig1}b-c]. Similar physics also appears in transition metal dichalcogenides MX$_2$. These examples show that the orbital response reflects topological properties of centrosymmetric materials and can be connected to experimentally observable quantities. %
 
\section*{Results}

\subsection{Model.}
We consider an electron model with four bands. The most general form of the two-band $4\times 4$ Bloch Hamiltonian is
	\be \label{genH}
	H = d_0(\bm k) I_4 + \sum_{a = 1}^5 d_a(\bm k) \Gamma^a,
	\ee
where $I_4$ is $4 \times 4$ identity matrix and $\Gamma^a$ ($a=1,\ldots,5$) are the Dirac matrices that satisfy the Clifford algebra $\left \lbrace \Gamma^a,\Gamma^b \right \rbrace = 2 I \delta^{a,b}$. We choose them as $ \Gamma^a = \left\lbrace \tau_z, \tau_y,  \tau_x \sigma_x , \tau_x  \sigma_y , \tau_x \sigma_z \right\rbrace$, where Pauli matrices $\sigma_\alpha$ and $\tau_\alpha$ ($\alpha=x,y,z$) are for spin and orbital degrees of freedom, respectively; $\tau_\alpha$ can be atomic orbitals, different layers in layered two-dimensional materials, or different sublattices.
The model~(\ref{genH}) has two doubly-degenerate bands $(1,1')$ and $(2,2')$ with energies $ \epsilon_{1}(\bm k) = \epsilon_{1'}(\bm k)$, $ \epsilon_{2}(\bm k) = \epsilon_{2'}(\bm k)$, respectively:
    \be
    \epsilon_{1,2}(\bm k) =d_0(\bm k) \mp \sqrt{\sum_{a=1}^5d_a(\bm k)},
    \ee
where the degeneracy is a consequence of the Kramers theorem in the presence of the inversion symmetry.

The time reversal and inversion symmetries are implemented by the operators
   \be \label{TR_P}
           {\mathcal T} = i \sigma_y \hat K,\qquad P = \tau_z {\mathcal P},
    \ee
 where $\hat K$ denotes complex conjugation and $\mathcal P$ is the inversion operator acting on the real-space coordinate ${\mathcal P}\,f(\bm r) = f(-\bm r)$. The  Hamiltonian~(\ref{genH}) is invariant under both ${\mathcal T}$ and $P$ operators, provided that functions $d_{0,1}(\bm k)$ [$d_{2,\ldots,5}(\bm k)$] are even [odd]  with respect to the inversion of momentum.

\subsection{Orbital current.} 
Systems with inversion symmetry allow for a nonzero second-order dc response for certain kinds of orbital currents; more specifically, these currents correspond to the transport of orbitals that transform into each other under inversion (See Materials and Methods for details). For concreteness, we consider the orbital current given by
   \be \label{staggered} 
   {\bm J}_\text{orb} 
    = 
      \left \langle {\bm u} \right \rangle
    =
     \left \langle \left\lbrace \tau_x, \bm v \right\rbrace \right \rangle,
    \ee
where $\vstag=\hat{P}^1_x \bm v \hat{P}^1_x - \hat{P}^2_x \bm v \hat{P}^2_x =  \left\lbrace \tau_x, \bm v \right\rbrace$ is the staggered velocity operator, $\hat P^{1,2}_x = \frac{1}{2}  \left(I_2\pm \tau_x \right)$ are the orbital projection operators, and $\langle\cdots\rangle$ is the thermal average. In contrast to the current of the orbital angular momentum~\cite{PhysRevLett.95.066601,xiao2020detection}, the operator $\bm u$ does not change sign under $P$. Thus, the corresponding second-order nonlinear optical response is allowed by the symmetry.
   
An example of the orbital degree of freedom corresponding to the current in Eq.~\eqref{staggered} is the electric polarization. Suppose $\tau_z=1$ is an $s$ orbital $\ket{s}$ and $\tau_z=-1$ is a $p$ orbital state $\ket{p_{\sigma}}=\ket{p_x+i\sigma p_y}$ ($\sigma = \pm$). In this case, the eigenstates of $\tau_x$ are $\frac1{\sqrt2}(\ket{s} \pm \ket{p_{\sigma}})$. In particular, this reflects the case for the BHZ model \cite{Bernevig1757}. Such a superposition will cause a displacement of the electron density in the $x$-direction, as is shown in Fig.~\ref{fig1}d. This displacement corresponds to the intracell electric polarization. The flow of the polarization can lead to an accumulation of the polarization at the edges of the sample or in the contact leads.
   
 We note that, in addition to the one introduced above, one can consider other components of orbital and spin currents. The $\tau_x$ orbital current operator is the simplest one that by construction involves  only orbital degrees of freedom in the definition and possesses the necessary symmetry properties.  In the Methods section we use symmetry considerations to classify all such operators and discuss their properties. We also show that the orbital current introduced above can be split into conserved and nonconserved parts, similarly in spirit to the Ref.~\cite{PhysRevLett.95.066601}.

\subsection{Second-order orbital response.}
We calculate the nonlinear orbital current by combining Floquet band formalism with the Keldysh Green's functions technique~\cite{Haug2007,Morimotoe1501524}, see the Methods section.
In the case of the linearly polarized light,  the orbital current reads $\bm J_\text{orb} =  - i \mathrm{Tr} \left( \bm u G^{<}\right)$, where $G^<$ is the lesser Green's  and the trace denotes the combination of integrating over the frequencies $\omega$, integrating over the Brillouin zone, and the matrix trace in the spin + orbital basis. For the model in (1) and assuming a 2D system, the calculation gives:
	\be \label {total_curr}
	\bm J_\text{orb} = \int \frac{d^2  \bm k}{(2 \pi)^2} \ \left( \bm j^{(1)}_\text{orb} + \bm j^{(2)}_\text{orb} + \bm j^{(3)}_\text{orb} \right),
	\ee

where
	\bea 
	\! \! \! \! \!  \! \! &\bm{j}^{(1)}_\text{orb}& =  i \frac{ \Gamma |A|^2  }{\Delta_\Omega^2 + \Gamma^2}   \Tr \left[ V^+  {\bm U}_k - ( {\bm U}_k)^+ V\right] , \label{res}\\
	\! \! \! \! \! &\bm{j}^{(2)}_\text{orb}& = 
	\frac{\left( \epsilon_2({\bm k}) - \epsilon_1({\bm k}) - \Omega \right)  |A|^2  }{\Delta_\Omega^2 + \Gamma^2} \Tr \left[ V^+  {\bm U}_k  + ( {\bm U}_k)^+ V\right]  \label{res2}  \\
    \! \! \! \! \!  \! \! &\bm{j}^{(3)}_\text{orb}&  =  \! {\bm u}_{11} \!  + \! 2 \tau \frac{|A|^2  \Gamma }{\Delta_\Omega^2 + \Gamma^2} \det \! V \! \left [  {\bm u}_{22} \!  - \! {\bm u}_{11}  \right ]\! + \!\left ( 1,2 \! \leftrightarrow \! 1' \! ,2' \right ) \!  \label{res3}
	\eea
Here  $\Delta_\Omega^2 = (\epsilon_2 - \epsilon_1 - \Omega)^2 + 4 |A|^2  \left( |v^0 _{12}|^2 + |v^0 _{1'2}|^2 \right)$, $\Omega$ is the light frequency, $\Gamma^{-1}$ is the decay time, and $\bm A$ is the vector potential that corresponds to the intensity of the light. The prefactor $\Gamma/(\Delta_\Omega^2 + \Gamma^2)$ becomes a delta function when $\Gamma\to 0$. The matrices $V$ and $ {\bm U}_k$ are the interband components of the velocity operator,
    \be \nonumber
    V = \begin{pmatrix}
	v^{0}_{12}  &  v^{0}_{12'}\\ 
	v^{0}_{1'2} &  v^{0}_{1'2'}
	\end{pmatrix},
	\ee
and the derivatives of the staggered velocity operator
    \be \nonumber
    {\bm U}_k = \begin{pmatrix}
	\left(\frac{\partial  {\bm u}}{ \partial k_\alpha} \right) _{12}  & 	 \left(\frac{\partial  {\bm u}}{ \partial k_\alpha} \right) _{12'}\\
	\left(\frac{\partial  {\bm u}}{ \partial k_\alpha} \right) _{1'2} & 	 \left(\frac{\partial  {\bm u}}{ \partial k_\alpha} \right) _{1'2'}
	\end{pmatrix},
    \ee
respectively. Here, $v^0 = \partial H/ \partial k_\alpha$ and $k_\alpha \equiv \bm k \cdot \bm A / A$.

The first term in Eq.~\eqref{total_curr} corresponds to  the shift-current-type response, which will be the main focus of our work. Equation~\eqref{res} has the same structure as the shift current in nonlinear response theory~\cite{Kraut1979,vBaltz1981}.  The current in Eq.~\eqref{res2} vanishes in the presence of time reversal symmetry, whereas the last contribution, Eq.~\eqref{res3}, is the injection current proportional to the relaxation time. The injection current vanishes for the case of linearly polarized light.  However, it gives a non-zero current in the case of circularly-polarized light; this will be discussed elsewhere. In the rest of this work, we study $\bm j^{(1)}_\text{orb}$ focusing on the case of linearly polarized light.

The presence of $U(2) \times U(2)$ gauge symmetry may lead to a non-Abelian Berry phase in the system.  It enters the expression for orbital shift current $\bm j^{(1)}_\text{orb}$ explicitly through covariant derivative in $k$-space $	\left(\frac{\partial  {\bm u}}{ \partial k_\alpha} \right) _{ij} = \frac{\partial  {\bm u}_{ij}}{ \partial k_\alpha} +  i [a^\alpha,\bm u]_{ij}$, where  the matrix elements for non-Abelian Berry connection for the quasimomentum in the direction of $\bm A$ are $a^\alpha_{i j} = - i \bra{i} \frac{\partial}{\partial k_\alpha} \ket {j}$.  For more details, we direct the reader to the Supplementary section.

\subsection{Orbital current in four-band model.} 
\begin{figure}[t] 
	\includegraphics[width= \columnwidth]{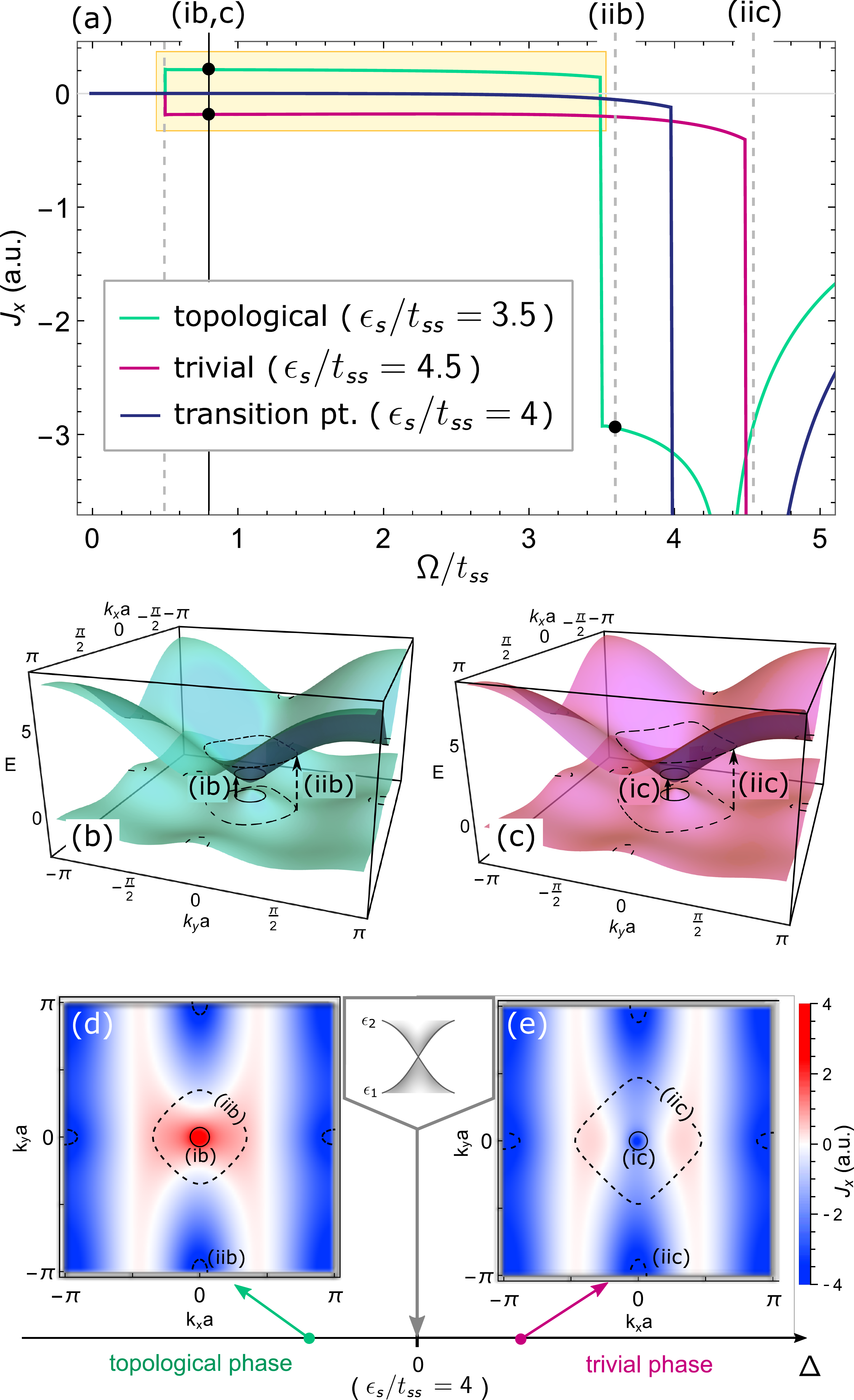}
	\caption{(a) The dependence of the total orbital current $J^{(1)}_{orb} \equiv J_x$ on the light frequency for BHZ model. $\epsilon_s/t_{ss} = 3.5$ is below   and $\epsilon_s/t_{ss} = 4.5$   is above the phase transition from topological to the trivial insulator that occurs as   $\epsilon_s/t_{ss} $ crosses the value 4.  The  dashed gray line corresponds to the minimum frequency. The bandstructure is shown below (b) and above (c) the topological phase transition where the lines $(ib,c)$ and $(iib,c)$  are the equal energy difference contours at which the optical transitions (shown by arrows) occur. The corresponding frequencies are indicated in panel (a).	(d), (e) - maps of the orbital current density $j_x(\bm k)$ at  (b) $\epsilon_s/t_{ss} = 3.5$  and (c) $\epsilon_s/t_{ss} = 4.5$.  The other parameters are  $\epsilon_p = t_{pp} = 0$, $t_{sp}/t_{ss} = 1$.  } \label{fig2}
	\end{figure}

For the four-band model in Eq.~(\ref{genH}),  the current parallel to the light polarization reads
	\be \label{curr_H_pars}
	\begin{split} 
	J^{(1)}_\text{orb} =\! \! \! \! \underset{\Delta_\Omega(\bm k) = 0}{\int} \! \frac{d k}{2 \pi}  
	 \frac{4 |A|^2}{ \epsilon |\nabla_k \Delta_\Omega|} \!
	\left[ d_0'' (d_2 d_1' \! - \!  d_1 d_2') + \varepsilon^{l sr} g_l g_s' g_r'' \right] \!   \! ,
	\end{split}
	\ee
where the primes denote derivatives over $k_A$, $\epsilon = \sqrt{\sum_{i = 1}^5 d_i^2}$, and $\varepsilon$ is the Levi-Civita tensor. We have introduced a short-hand notation for the spin-orbit coupling coefficients $\bm g (\bm k) = \lbrace d_3,d_4,d_5\rbrace$. The integral is over the surface in the Brillouin zone that satisfies $\Delta_\Omega=0$ [Fig.~\ref{fig2}b-c]. For a direct-gap semiconductor  when $\Omega$ is close to the band gap, the integral is taken over the boundary of a small pocket around the band bottom (e.g. contour (ib) in Fig.~\ref{fig2}b). This is the case for topological insulators in the vicinity of the topological phase transition.

To uncover the relation between the topological index and $j^{(1)}_\text{orb}$, we consider a situation where  $d_1(\bm k)$ smoothly changes sign at an odd number of time-reversal-invariant momenta (TRIM) $\bm \Gamma_i$ ($i=0, 1, 2,\dots$). Due to this property $d_{2,3,4,5}(\bm k)$ vanish at $\bm\Gamma_i$, but their first derivatives remain nonzero. Therefore, Eq.~\ref{curr_H_pars} for $\Omega \approx \epsilon_2(\bm k=\bm\Gamma_i)-\epsilon_1(\bm k=\bm\Gamma_i)$ reads 
	\be 
	\begin{split} 
	j^{(1)}_\text{orb} (\bm k \approx \bm \Gamma_i) \propto	- d_0'' d_2' \textrm{sign}(d_1)_{\bm k = \bm \Gamma_i}  ,\label{j1_topo_index}
	\end{split}
	\ee
where we have used $\epsilon(\bm \Gamma_i) = |d_1(\bm \Gamma_i)|$.

The signs  of $d_1(\bm k)$ at at $\bm{\Gamma}_i$-points can be used to express the $\mathbb{Z}_2$ invariant of Hamiltonian~\eqref{genH} as
   \be
    \delta_{\mathbb Z_2} = \prod_i \frac{\sqrt{\mathrm{Det}[B(\bm \Gamma_i)]}}{\mathrm{Pf}[B(\bm \Gamma_i)]} = \prod_i \text{sign} [d_1 (\bm k =\bm \Gamma_i)],
    \ee  
where $B_{mn}(\bm k)= \bra{u_m (-\bm k)} {\mathcal{T}} {\ket{u_n (\bm k) }}$ is the sewing matrix\cite{PhysRevB.76.045302}. The $\mathbb{Z}_2$ indicator $\delta_{\mathbb Z_2}$ corresponds to the $\mathbb{Z}_2$ topological index in 2D or to the strong topological number in 3D if the sum is over all TRIMs in the half of the Brillouin zone (BZ). If the summation runs over TRIMs in the same plane in a half of BZ, the index becomes a weak topological number in 3D. Comparing the $\mathbb{Z}_2$ index to Eq.~\eqref{j1_topo_index}, we obtain that
	\be 
	\begin{split} 
	\prod_i \textrm{sign} \left [ j^{(1)}_\text{orb}(\bm k \approx \bm \Gamma_i) \right ] \propto  \delta_{\mathbb Z_2}.
	\end{split}
	\ee
This links the behavior of a physical observable, the orbital current, to the indicator of topological phase transition.

The quantity defined above changes the sign when the $\mathbb{Z}_2$ index changes as $d_0''$ and $d_2'$ are finite at the phase transition. In many relevant examples, the gap closes at one TRIM, therefore measuring the sign of the orbital current when inducing transitions near that specific point would yield the topological index. Note that neither $d_0''$ nor $d_2'$ vanishes at this point unless it is required by some additional symmetry. Therefore, a smooth change of the sign as $d_1 (\bm \Gamma_i)$ leads to a change in the sign of the orbital current, which is a robust effect regardless of the material. 

 One can see that the sensitivity of the orbital current to the change of the topological phase is physically meaningful from the following considerations. At the phase transition the change of the parity of the wavefunctions occurs at the corresponding TRIMs\cite{PhysRevB.74.195312}. Under crossing the phase transition, the wavefunctions undergo the following  change:
	\be \label{wfs}
	\begin{split}
	&\psi_1 (\bm k)\big|_{\Delta} = - P \psi_2 (-\bm k)\big|_{-\Delta}, \\
	&\psi_{1'} (\bm k)\big|_{\Delta} = - P \psi_{2'} (-\bm k)\big|_{-\Delta},
\end{split}
	\ee
where $\Delta = 2 d_1(\bm k = 0)$ is value of the bandgap that accounts for the possibility of band inversion due to the change in the sign of $d_1$. Equation~\eqref{wfs} implies that the wavefunctions above the transition are related to another band's parity-inverted wavefunctions below the transition. We use that $ P \psi_2  =  - {\mathcal T} \psi_{2'}$ and $ P \psi_{2'}  =  {\mathcal T} \psi_{2}$ to relate the action of parity and  time-reversal symmetries at the same point $\bm k$ in the BZ. In simple terms, flipping of the  eigenfunction parity reverses the arrow of time for transition processes, thus changing the sign of the current. Let us use Eq.~\eqref{wfs} to argue that the expression for the orbital current in Eq.~\eqref{res} changes sign. Both the $\bm v$ and $\bm v^z$ velocity operators that enter the expression for the orbital current density are odd under $\mathcal T$. Using this fact and the above transformation of the wavefunctions we obtain $v_{12}  (\bm k)\leftrightarrow v^*_{1'2'} (- \bm k)$,  $v_{12'} (\bm k) \leftrightarrow -v^*_{12'} (-\bm k)$ and $v_{1'2} (\bm k) \leftrightarrow - v^*_{1'2} (-\bm k)$.
It follows that ${\rm Tr}\left[V^+ \widetilde{\bm U}_k\right]$ becomes complex conjugated at the point $-\bm k$ upon crossing the  phase transition. Since the expression for the shift current in Eq.~\eqref{res} is the imaginary part of the trace and only the $\bm k$-even part of the current density survives after  integration, the shift current must necessarily change sign. Thus, the orbital current probes the change in the parity of wavefunctions at the topological phase transition. 

A similar argument explains why the current vanishes near the gap-closing TRIM exactly at the phase transition point. At this point the eigenfunctions of the Hamiltonian  also become parity eigenstates. In this case, there is no difference between  electron density distributions in the initial and final states, and thus, there can be no shift orbital current. 

\subsection{Bernevig-Hughes-Zhang model.}
As an example, we  consider Bernevig-Hughes-Zhang (BHZ) model on a square lattice,~\cite{Bernevig1757}  an effective model for HgTe/CdTe quantum wells. The model corresponds to the Hamiltonian in Eq.~\eqref{genH} with the wave function basis $\ket {s,\uparrow}$, $\ket {s,\downarrow}$,  $\ket {p_x + i p_y,\uparrow}$, $\ket {p_x - i p_y,\downarrow}$ and $d_i(\bm k)$ shown in Table~\ref{t1} \cite{PhysRevB.76.045302}. By varying the thickness of the quantum wells, the constants $\epsilon_s$ and $\epsilon_p$ can be tuned to the transition point between a trivial and topological insulator~\cite{Bernevig1757}. Using Eq.~\eqref{curr_H_pars}, the orbital current density in the direction of light polarization reads:
	\be \label{Jbhz}
	j^{(1)}_\text{orb} = 	\frac{ 2\Gamma |A|^2  }{\Delta_\Omega^2 + \Gamma^2}  \frac{4 a^3 t_{sp} \left( t_{ss}- t_{pp} \right) }{\epsilon} f(k_x,k_y),
	\ee
with $f(k_x,k_y) = - \cos  k_x a [ (\epsilon_p - \epsilon_s) \cos  k_x a + 2 (t_{ss} + t_{pp})(1 + \cos  k_x a \cos  k_y a )]$, and $a$ being the lattice constant.  In the vicinity of the center of the BZ, the current along $x$-direction reads
\be
j^{(1)}_\text{orb} \big|_{\Gamma\rightarrow 0} \approx	  - 8 \pi a^3 |A|^2 \delta(\Delta_\Omega) t_{sp} (t_{ss} - t_{pp})  \, \text{sign}\left[ d_1(0)\right].
\ee

		\begin{table} [!b]
		\begin{tabular}{ p{0.9cm}|c  }
			\hline \hline
			$d_0$ & $(\epsilon_s + \epsilon_p)/2 - (t_{ss}- t_{pp})(\cos \bm k \cdot \bm a_1+\cos \bm k \cdot \bm a_2)$ \\
			$d_1$ & $(\epsilon_s - \epsilon_p)/2 - (t_{ss}+ 	 t_{pp})(\cos \bm k \cdot \bm a_1+\cos \bm k \cdot \bm a_2)$ \\
			$d_2$ & $2  t_{sp}\sin  \bm k \cdot \bm a_1$ \\
			$d_3, d_4$ & 0 \\
			$d_5$ & $2  t_{sp}\sin  \bm k \cdot \bm a_2$ \\
			\hline \hline
		\end{tabular}
		\caption{Coefficients of the BHZ Hamiltonian  from Ref.~[\onlinecite{PhysRevB.76.045302}]. }  \label{t1}
	\end{table}

Equation~\eqref{Jbhz} explicitly shows the sign change at $d_1(\bm k=\bm 0)=\{\epsilon_s - \epsilon_p - 4 (t_{ss} + t_{pp})\}/2=0$, reflecting the topological phase transition associated with the gap closing at the $\Gamma_0$ point. 
In Fig.~\ref{fig2}a, we plot the total orbital shift current as a function of the light frequency below and above the phase transition. At small frequencies the light probes the current density near the $\bm \Gamma_0$ point, where the gap closing occurs at the topological phase transition.  At low frequencies, the sign of the current in Fig.~\ref{fig2}a is the opposite below and above the topological phase transition (the relevant area is highlighted in yellow). The value of the current remains nearly constant when the surface of the integral is the boundary of a pocket. The difference in the sign of the current density, that is indicative of the topological phase transition, is also shown in Fig.~\ref{fig2}d-e.

Exactly at the transition point, the current vanishes in $\bm k \cdot \bm p$ approximation, which predicts the behavior of the current at small frequencies. For the BHZ model in this section, the $\bm k \cdot \bm p$  Hamiltonian has the form
\be \label{Hkp}
H_{\bm k \cdot \bm p} = c_0 k^2 +  \delta \tau_z + 2 t_{sp} a \left ( k_x \, \tau_y + k_y \, \tau_x \sigma_z \right ) .
\ee
Here we neglected a constant contribution,  the mass term is $\delta = d_1(\bm k = \bm0)$ and the coefficient $c_0$ comes from the series expansion of $d_0(\bm k)$. The Hamiltonian in \eqref{Hkp} produces an orbital current  $j_{\text{orb}} \propto \delta$, which equals to zero at the phase transition where $\delta = 0$. In our particular example [Fig.~\ref{fig2}a, $\epsilon_s/t_{ss}=4$] it acquires a correction only at quite large frequencies, greater than $\sim 2.5 t_{ss}$.   

On the other hand, the magnitude and the sign of current behaves distinctly for high frequencies, when the vicinity of other TRIMs contributes to the current. At these frequencies the topology of the equal energy-difference contours (shown in Fig.~\ref{fig2}b-c and d-e) changes as the light frequency increases. At $\Omega \ge \epsilon_s - \epsilon_p$ the contour encircles TRMs $(0,\pi)$ and $(\pi,0)$, in addition to $(0,0)$. This corresponds to contours $(iib)$ and $(iic)$ in Figs. \ref{fig2}b-e, causing a jump of the sign of the current as the current density near these points is the same as that near $\bm \Gamma_0$ but is opposite in sign, whereas the JDOS is finite regardless of the radius of the circle, as is expected in two dimensions. The contour encounters van Hove singularities upon further increase the frequency, which manifests as a growth of the response.

In experiment, the orbital current can give rise to an electric polarization at the edges of the material. In the BHZ model, the average of the operator $\tau_x$ is proportional to the intracell polarization in $x$ direction, $\hat P_x = -({e}/{V}) \hat x$ (see Methods) where $V$  is the volume of the unit cell. Therefore, the  orbital polarization due to the photo-induced orbital current is  related to the electrical intracell polarization [Fig.~\ref{fig1}c-d]. The spatial distribution of the polarization is observable using, for example, optical techniques~\cite{Denev2011} that can be extended to the photo-induced orbital current as well.

For an estimate of the magnitude of the current in HdTe/CdTe quantum wells, we use $\epsilon_s = 3.99$ eV, $\epsilon_p = t_{pp} = 0$, $t_{sp} =t_{ss}$=1 eV, which corresponds to a bandgap of 10 meV. Eq.~\eqref{curr_H_pars} can be further simplified to 
	\be \label{eq14}
	J^{(1)}_\text{orb} = - \frac{4 e^2}{  \hbar^3 \Omega^2} \frac{I}{c \varkappa \epsilon_0}  \underset{\Delta_\Omega(\bm k) = 0}{\int} \frac{d k}{2 \pi}  
	 \frac{1}{ \epsilon |\nabla_k \Delta_\Omega|}
	 d_0''   d_2' d_1 .
	\ee
Here, we restored the constants $e$, $\hbar$ and $c$. The light intensity is $I = \epsilon_0 \varkappa  c  |\bm E|^2$, and $\varkappa$ is the effective dielectric permittivity.
Plugging the above parameters  for the effective  model for HdTe/CdTe quantum wells into Eq.~\eqref{eq14} and assuming $\epsilon = 10$, $I=$ 10 GW cm$^{-2}$, and that the frequency $\Omega$ equals to the bandgap,  we obtain $J_\text{orb} \approx 10^{24}$ m$^{-1}$ s$^{-1}$.  An intensity equal to 10 GW/cm$^2$ was chosen because such values recently became achievable in THz experiments~\cite{Hirori2011THz,Sato2013THz,Liu2017THz}.

\subsection{ 1T' topological phase of MX$_2$.}
Another example of 2D TI is  transition metal dichalcogenides MX$_2$ (where M stands for a metal and X is a chalcogen atom) in 1T' phase,~\cite{biswapriyo2020tuneable,PhysRevB.93.125109} such as MoS$_2$, WSe$_2$ and WTe$_2$. The Hamiltonian of MX$_2$ in 1T' phase is well described by the four-band  Hamiltonian in Eq.~(\ref{genH}).~\cite{qian2014quantum,liu2015assessment} In $\bm k\cdot\bm p$ approximation, the coefficients of the Hamiltonian are given in Table~\ref{t2}.
In MoS$_2$, the effective masses for $p_x$ ($p_y$)orbitals are\cite{biswapriyo2020tuneable} $m^p_{x} = 0.48 m_e$ ($m^p_{y} = 0.29 m_e$), and $m^d_{x(y)} = 2.32 (0.92) m_e$ for the two $d$-orbitals. $ \delta = -0.27$~eV corresponds to $d-p$ band inversion, and velocities are $v_1 = 0.23\times 10^5$ m/s and $v_2 = 3.38 \times 10^5 $ m/s. Plugging the numbers into Eq.~\eqref{eq14} and performing the integral, for the light frequency equal to the band gap at $\Gamma_0$-point ($\Omega = 2| \delta|$), the order of magnitude estimation of the orbital current density is $J_{orb} \approx 10^{22}$ m$^{-1}$ s$^{-1}$. Here, we assumed $I=10$  GW cm$^{-2}$ and the dielectric constant of $\varkappa = 10$.
		\begin{table} [t]
		\begin{tabular}{ p{0.9cm}|c }
			\hline \hline
			$d_0$ & $\sum_{i = x,y} \frac{\hbar^2 k_i^2 }{4 }\left (\frac{1}{m_i^d} - \frac{1}{m_i^p} \right )$ \\
			$d_1$ & $-  \delta -\sum_{i = x,y} \frac{\hbar^2 k_i^2 }{4}\left (\frac{1}{m_i^d} + \frac{1}{m_i^p} \right )$ \\
			$d_2$ & $-v_2 \hbar k_x$ \\
			$d_3$ & $v_1 \hbar k_y$ \\
			$d_4, d_5$ & $0$ \\
			\hline \hline
		\end{tabular}
		\caption{Coefficients of the BHZ Hamiltonian  for MX$_2$ family near the $\bm \Gamma_0$-point~\cite{qian2014quantum,liu2015assessment}. $m_{x,y}^{d,p}$ are the effective masses of $d$ and $p$- bands along different directions, respectively, $v_{1,2}$ are the velocities and $\delta$ is the value of the band inversion.}  \label{t2}
	\end{table}

\section*{Discussion}

In this work, we studied the optical dc response of an newly constructed orbital current in time-reversal and inversion symmetric materials. We find that the sign of orbital current reflects the $\mathbb{Z}_2$ index of centrosymmetric topological insulators, which makes the orbital current an experimental probe for the topological properties of electronic bands. At the topological phase transition, the orbital current exhibits a robust sign change. We illustrated these properties by considering the BHZ Hamiltonian and an effective model for MoS$_2$. In these materials, our estimate for the orbital current gives $10^{22}-10^{24}$ m$^{-1}$ s$^{-1}$ for a light power of 10 GW cm$^{-2}$.

The orbital current generated in a semiconductor accumulates intracell electric polarization as schematically shown in Fig.~\ref{fig1}.  We expect that the accumulated polarization decays over timescales on the order of the orbital momentum relaxation time $\tau \approx 10^{-14}-10^{-15}$ s, which is determined by the value of the bandgap (see discussion in the Materials and Methods section). Taking the relaxation into account, the number of  carriers with $\langle \tau_x \rangle = 1$ that can be transported  per unit cross-section length and subsequently accumulated is roughly equal to $\tau J_\text{orb}$.  The dipole moment of a fully orbitally polarized  unit cell ($\langle \tau_x \rangle = 1$) equals  $10^{-29}$ C m, and an orbital current of magnitude $10^{22}-10^{24}$ m$^{-1}$ s$^{-1}$ produces $\tau J_\text{orb}$ on the order of $10^{10}-10^{7}$ m$^{-1}$. The accumulated dipole moment per meter of cross-section length therefore equals  $10^{-19}-10^{-22}$~C.  

A similar effect may also take place in other kinds of systems. For instance, by applying time-periodic driving to a BHZ model, one can induce a phase transition to a Floquet topological insulator state  \cite{lindner2011floquet}. It was shown that even when the driving potential breaks time-reversal symmetry, the system is still described by the BHZ model with different effective parameters when one works in an appropriate rotating frame of reference. We expect the orbital current to be a suitable probe for such a phase transition. Application of optically-driven orbital currents to the detection of other topological phases and strongly correlated states is another interesting problem that remains an open question.

\section*{Acknowledgements}
We are grateful to Takahiro Morimoto and Zhanybek Alpichshev for fruitful discussions.  MD was supported by Austrian Agency for
International Cooperation in Education and Research (OeAD-GmbH) and by the John Seo Fellowship at MIT. HI was supported by JSPS KAKENHI Grant Numbers JP19K14649 and JP18H03676, and by UTokyo Global Activity Support Program for Young Researchers.

\section*{Materials and Methods}
\setcounter{equation}{0}
\renewcommand\theequation{M\arabic{equation}}

\subsection{Derivation of nonlinear shift current.}
The coefficients $d_{0,1,2,3,4,5}(\bm k)$  of the Hamiltonian \eqref{genH} have to be either even or odd in the BZ:
    \be \label{d_parity}
	\begin{split}
	&d_{0,1} (\bm k) = d_{0,1} (- \bm k), \\
	&d_{2,3,4,5} (\bm k) = -d_{2,3,4,5} (- \bm k). \\
	\end{split}
	\ee
The energy eigenvalues are 
	\be 
	\epsilon_{1,2} (\bm k)= d_0(\bm k) \pm \sqrt{|d_1 (\bm k)|^2 + \sum_{i=2}^5 |d_i (\bm k)|^2 } = d_0(\bm k) \pm \epsilon(\bm k).
	\ee
The term $d_1(\bm k)$ has special place in this model because $\Gamma_1$ is the only of five matrices that is even under $\hat P$ and $\hat {\mathcal T}$. At TR-invariant momenta $d_{2,3,4,5}$ all vanish and the parameter  $d_1(\bm k)$ represents the mass term that governs the topological phase transition.  When it changes sign at odd number of points, this changes  the (strong) topological index in 2D (3D).

To derive the expression for the shift current, we use the Floquet bands formalism combined with Keldysh Green's functions \cite{Morimotoe1501524}.   We choose the wave functions  1 and 1'  (2 and  2') to be the $\hat {\mathcal T} \hat P$-partners, which partially fixes the $U(2)\times U(2)$ gauge redundancy.
	
Let us write the vector potential $\bm A(t) = i \bm A e^{i \Omega t} - i \bm A^* e^{-i \Omega t}$ and assume that incident electromagnetic wave is linearly polarized. We consider the Floquet bands with indices $-1$ (valence bands) and $0$ (conduction bands).   The Floquet Hamiltonian reads:
	\be
	H^F = \begin{pmatrix}
	\epsilon_1 + \Omega & 0 & -i A^*v^{0}_{12}  & -i A^*v^{0}_{12'}\\ 
	0 & \epsilon_1 + \Omega & -i A^*v^{0}_{1'2} & -i A^*v^{0}_{1'2'}\\ 
	i A v^{0}_{21} & i A v^{0}_{21'} & \epsilon_2  & 0 \\ 
	i A v^{0}_{2'1} & i A v^{0}_{2'1'} & 0 & \epsilon_2 
	\end{pmatrix} 
	\ee
where $v^0 = \frac{\partial H_0(k)}{\partial \bm k} \cdot \bm A/A$ is the velocity along the polarization direction. The dc orbital current is given by
	\be
	\bm J_\alpha = -i\int \frac{d \omega}{2 \pi}\int d  k \   \mathrm{Tr} \left( \bm v^{F}_\alpha G^{<}\right),
	\ee
where $\bm v^{F}_\alpha$ is the orbital velocity operator defined in the previous section taken in Floquet representation. The lesser Green's function of the Floquet Hamiltonian is $G^{<}(\omega) = G^R (\omega)  \Sigma ^{<} G^A (\omega)$, 
where $\Sigma ^{<}$ is the self-energy. 
In the gauge that demands  the spin components within the pairs $1,2$ and $1',2'$ to be parallel, the self-energy acquires diagonal form $\Sigma ^{<} = i \Gamma \frac{1}{2} \left( I_2 + s_z \right) $, where $s_z$ is a Pauli matrix that acts on the band degree of freedom. 

Evaluating the integral over $\omega $ and collecting the terms with the similar structure, we arrive at Eq.~\eqref{total_curr}.

\subsection{Connection between orbital degrees of freedom and intracell polarization.}
In BHZ model the basis consists of states  $\ket {s,\uparrow}$, $\ket {s,\downarrow}$,  $\ket {p_x + i p_y,\uparrow}$ and $\ket {p_x - i p_y,\downarrow}$. Let us consider a simplified argument and assume that the Wannier functions for $p$-orbitals have the form $w_{p_x \pm i p_y } (\bm r) = c (x \pm i y) w_s (\bm r)$, where $c$ is a constant that is assumed to be real.   The matrix  of the $x$-component of the electrical polarization operator is
\begin{widetext}
\be \label{Px}
	\begin{split} 
	P_x  =  - \frac{e}{V_2} \int_\text{UC} d\bm r\begin{pmatrix}
      I_2 & x  I_2 + i  y \sigma_z   \\ 
    x I_2 -i  y\sigma_z     & \left(x^2 +  y^2  \right )  I_2
    \end{pmatrix} c x  |w_s(\bm r)|^2
     = a \tau_x  \otimes I_2,
	\end{split}
\ee
\end{widetext}
where integral is performed over unit cell, $V_2$ is the area of 2D unit cell, and $a$ is a real constant obtained after integration $a = - \frac{e}{V_2}  \int d\bm r c x^2 |w_s(\bm r)|^2$ and $2\times 2$ matrix $I_2$ acts on spins. In a more general case, expression for $P_x$ will contain a linear combination of matrices $\tau_{x,y}$ with complex coefficients. However, by applying an appropriate rotation  we retrieve Eq.~\eqref{Px} again, confirming that this is a general result.

\subsection{Symmetry-based classification of allowed responses for two-dimensional layer groups and three-dimensional  centrosymmetric groups.}
Let us consider a second-order response in the current associated with some operator $\mathcal O$. Phenomenologically, the dc response is : 
    \be \label{symm1}
    j_i (\mathcal O) =   \sigma_{i j k} (\mathcal O) E_j E^*_k,
    \ee
where $\sigma_{i j k} (\mathcal O) $ can be a third rank tensor is $\mathcal O$ is scalar (pseudoscalar) or a highter-rank tensor if the operator  $\mathcal O$  has tensor structure. We assume that $j_i(\mathcal O)$ transforms under a representation $\Gamma( v_i \times \mathcal O ) \sim \Gamma(vec) \otimes \Gamma (\mathcal O)$. Because we are considering centrosymmetric materials, for the response \eqref{symm1} to be nonzero, $\mathcal O$ has to transform under an inversion-odd representation. 

Assuming that $\mathcal O$ transforms under an irreducible representation (irrep) of the corresponding point group, one can deduce the amount of independent response coefficients entering the tensor $\sigma_{i j k} (\mathcal O) $ that is symmetric in the last two indices as in the case of shift current. In Tables  \ref{tS1} and \ref{tS2} we summarize the results of the symmetry-based analysis for 2D and 3D centrosymmetric groups, respectively. 
Naturally, only inversion-odd representations of $\mathcal O$ give rise to nonzero response. The second columns of Tables  \ref{tS1} and \ref{tS2} can serve as guidance for choosing the appropriate observables for each symmetry group. Consider, for example, group $C_{6h}$. The irrep $A_u$ may correspond to $p_z$-orbital or a sublayer degree of freedom perpendicular to $z$-direction, $E_{1u}$ is a two-dimensional representation of that corresponds to $p_x \pm i p_y$ orbitals, and $B_u$ and $E_{2u}$ comprise parts of an electric octupole   \cite{hayami2018classification}.

\begin{table}[!t]
\begin{tabular}{c|l|l}
 \hline
                     Point   &  Representations of $\mathcal O$   &  $N_{tot}$\\ 
                     group  &   &  \\ \hline
$C_i$ &  $A_u$(12) & 12 \\
$C_{2h}$  & $A_{u}$(4); $B_{u}$(8) & 12 \\
$D_{2h}$ & $A_{u}$(2); $B_{1u}$(2); $B_{2u}$(4); $B_{3u}$(4) & 12\\
$C_{4h}$ & $A_{1u}$(2); $A_{2u}$(2); $E_{u}$(8) & 12\\
$D_{4h}$ &  $A_{1u}$(1); $A_{2u}$(1); $B_{1u}$(1); $B_{2u}$(1); $E_{u}$(4) & 8 \\
$S_{6}$ &  $A_{u}$(4); $E_{u}$(8) & 12\\
$D_{3d}$ &  $A_{1u}$(2); $A_{2u}$(2); $E_{u}$(4) & 8\\
$C_{6h}$ &  $A_{u}$(2); $B_{u}$(2); $E_{1u}$(6); $E_{2u}$(2) & 12\\
$D_{6h}$ &   $A_{1u}$(1); $A_{2u}$(1); $B_{1u}$(1); $B_{2u}$(1); $E_{1u}$(3); $E_{2u}$(1) & 8\\ \hline
\end{tabular}
\caption{Independent invariant tensor coefficients for  2D centrosymmetric point  groups.  The second column gives the irreducible representation for $\mathcal O$ where the response is allowed by symmetry and the number of independent tensor coefficients corresponding to a given representation. The third column provides the total possible number of nonzero tensor coefficients $N_{tot}$.  }  \label{tS1}
\end{table}

\subsection{Conservation of the orbital current.} \label{kp}
 The  orbital polarization operator $\tau_x$ can be separated into conserved ($\tau_x^{(c)}$) and nonconserved ($\tau_z^{(n)}$) parts, where conserved part obeys continuity equation and can be used to define a conserved current. By analogy with Refs.~[\onlinecite{Murakami2004,PhysRevLett.95.066601}], we write  $\tau_x^{(c)} = P^{v}\tau_x P^v+P^{c}\tau_x P^c$,  where $P^{v,c}$ are the projection operators on the subspaces of states belonging to  valence  and conduction bands, respectively. The non-conserved part is the difference between the full operator and its conserved component, $\tau_x^{(n)} = \tau_x - \tau_x^{(c)}$. We find that the second term in Eq.~\eqref{curr_H_pars} corresponds to the conserved orbital current, while the first term is the non-conserved part. 

To examine the dynamics of the non-conserved part of orbital charge, we find the equation of motion for the orbital polarization density $T = \psi^\dagger  \tau^x \psi$ to obtain: 
   \be \label{cont}
        \frac{\partial}{\partial t} \bm T - \nabla  \bm J_\text{orb} + \bm B= \bm Q.
    \ee
Here $\bm J_\alpha = \frac{1}{2}\psi^\dagger \left \{ \tau_\alpha, \bm v \right \} \psi $ is the orbital current associated with the flow of $\alpha$-th orbital  and $\bm Q =  \bm T \times \bm F$, $\bm F = \left \lbrace 0,0,\Delta \right \rbrace$, is the torque acting on the orbital polarization. The mass term $\Delta= 2 d_1(\bm k = 0)$ corresponds to the value of the gap at $\bm \Gamma_0$-point.  The third term in the expression is $B_\alpha = \psi^\dagger  i \bm \nabla~\cdot~\bm b [\tau_\alpha,\tau_y] \psi$, where $\bm b = \p d_2/ \p \bm k \big|_{\bm k = \bm \Gamma_i}$ and we are considering the continuous limit in the vicinity of $\Gamma_i$. The last term in the LHS of Eq.~\eqref{cont} couples $\nabla T_z$ to dynamics of $T_x$ and vice versa.  In a nearly-homogeneous setup this term is negligible.

Lastly, let us discuss the torque term  $\bm Q$. Near the topological phase transition the gap closes ($\Delta \rightarrow 0$), thus by  being close enough to the transition allows making the torque $\bm Q$  arbitrarily small. Away from the transition, the torque can be viewed as determining the rate of the dissipation of the orbital polarization in the material. 

\begin{table}[h]
\begin{tabular}{c|l|l}
\hline
                                         Point   &  Representations of $\mathcal O$   &  $N_{tot}$\\ 
                     group  &   &  \\ \hline
$C_i$ & $A_u$(18) & 18 \\
$C_{2h}$  & $A_u$(8); $B_u$(10) & 18\\
$D_{2h}$ & $A_u$(3); $B_{1u}$(5); $B_{2u}$(5); $B_{3u}$(5) & 18\\
$C_{4h}$ &  $A_u$(4); $B_u$(4); $E_u$(10) & 18\\
$D_{4h}$ &  $A_{1u}$(1); $A_{2u}$(3); $B_{1u}$(2); $B_{2u}$(2); $E_u$(5) & 13\\
$S_{6}$  & $A_{u}$(6); $E_{u}$(12) & 18\\
$D_{3d}$ & $A_{1u}$(2); $A_{2u}$(4); $E_{u}$(6) & 12\\
$C_{6h}$ & $A_{u}$(4); $B_{u}$(2); $E_{1u}$(8); $E_{2u}$(4) & 18\\
$D_{6h}$  & $A_{1u}$(1); $A_{2u}$(3); $B_{1u}$(1); $B_{2u}$(1); $E_{1u}$(4); $E_{2u}$(1) & 12\\
$T_{h}$ & $A_{u}$(1); $E_u$(2); $T_{u}$(5) & 8\\
$O_{h}$   & $A_{1u}$(1); $E_{u}$(1); $T_{1u}$(3); $T_{2u}$(2) & 7\\ \hline
\end{tabular}
\caption{Same as Table ~\ref{tS1} but for centrosymmetric groups in three dimensions.}  \label{tS2}
\end{table}

%

\clearpage
\onecolumngrid
\appendix
\section*{Supplementary}

\setcounter{equation}{0}
\renewcommand\theequation{S\arabic{equation}}

\subsection{Symmetries and the velocity operator}
	
The time-reversal operator 	is an antiunitary operator, thus the adjoint of it is defined as:
	\begin{equation}
	\bra{ \phi}\ket{\hat{T}\psi}  =  \overline{\bra*{ \hat{T}^\dagger \phi}\ket*{\psi}}.
	\end{equation}
 In the presence of both TR- and $P$- symmetries, one can choose the wavefunctions to be $\hat T \hat I$-partners and we denote the bands as 1, $1'$, 2 and $2'$, or explicitly:
	\begin{equation}
	\begin{split}
	\ket{\psi_{1'}}  &= \hat T \hat I \ket{\psi_{1}},\ \ket{\psi_{1}}  = - \hat T \hat I \ket{\psi_{1'}}, \\
	\ket{\psi_{2'}}  &= \hat T \hat I \ket{\psi_{2}},\ \ket{\psi_{2}}  = - \hat T \hat I \ket{\psi_{2'}}.
	\end{split}
	\end{equation}
Now we want to derive a relation between the matrix elements $\bra{\psi} v \ket{\psi}$ and $\bra{ \hat T \hat I \psi} v \ket{ \hat T \hat I \psi}$; it was also used that operator $v$ is invariant under $ \hat T \hat I$ symmetry.
	\begin{equation}
	\bra{ \hat T \hat I \phi} v \ket{ \hat T \hat I \psi}  =  \bra*{ \left(\hat T \hat I \right) ^{\dagger}  v  \left(\hat T \hat I \right) \psi}\ket*{  \phi} = \bra*{  \psi}  \left(\hat T \hat I \right)    v   \left(\hat T \hat I \right)^{\dagger}  \ket{  \phi}  = \bra*{  \psi}      v    \ket{  \phi}.
	\end{equation}
The last statement leads to the following relations:
	\begin{equation}
	v_{12} = v_{1'2'}^*, \ v_{1'2} = -v_{12'}^*, \ v_{11} = v_{1'1'}, \ v_{22} = v_{2'2'}.
	\end{equation}
Another property of the velocity matrix, which is not related to  the time-reversal, is that:
	\begin{equation}
	v_{11'} = v_{22'} = 0.
	\end{equation}
	The proof goes as follows
	\begin{equation}
	\begin{split}
	&H \ket{\psi_1 }  = \epsilon_1 \ket{\psi_1 }    \\
	&\frac{\partial H}{\partial k}\ket{\psi_1 }  + H \frac{\partial \ket{\psi_1 } }{\partial k}= \frac{\partial \epsilon_1}{\partial k} \ket{\psi_1 }  + \epsilon_1 \frac{\partial \ket{\psi_1 }    }{\partial k}    \\
	&\bra{\psi_{1'}} \frac{\partial H}{\partial k}\ket{\psi_1 }   + \psi_{1'} H \frac{\partial \ket{\psi_1 } }{\partial k}= \bra{\psi_{1'}}\frac{\partial \epsilon_1}{\partial k} \ket{\psi_1 }  + \epsilon_1 \psi_{1'}\frac{\partial \ket{\psi_1 }    }{\partial k} \\
	&v_{1'1} + \epsilon_{1'} \bra{\psi_{1'}}\ket {\frac{\partial \psi_1   }{\partial k}} = \frac{\partial \epsilon_1}{\partial k} \bra{\psi_{1'}}\ket{\psi_1 }  + \epsilon_{1}\bra{\psi_{1'}}\ket {\frac{\partial \psi_1   }{\partial k}} \\
	&v_{1'1}  = \frac{\partial \epsilon_1}{\partial k} \bra{\psi_{1'}}\ket{\psi_1 }  + \left (\epsilon_{1}  - \epsilon_{1'} \right )\bra{\psi_{1'}}\ket {\frac{\partial \psi_1   }{\partial k}} =  -i \left (\epsilon_{1}  - \epsilon_{1'} \right ) a_{1' 1} = 0.
	\end{split}
	\end{equation}

\subsection{Relation between the second-order response and Berry connection}

The expression for the shift current $\bm j^{(1)}_{\text{orb}}$ in Eq.~\eqref{res} can be related to  the non-Abelian Berry connections  $a_{\ell j} = - i \bra{\ell} \frac{\partial}{\partial k} \ket {j}$:
	
	\begin{equation} \label{res_a}
	\begin{split} 
	\bm j^{(1)}_{\text{orb}} &=   2\frac{\Gamma |A|^2  }{\Delta_\Omega^2 + \Gamma^2} \left\lbrace  \Re\left[ v^0_{21} {\bm u}_{12}\right]  \left( \frac{\partial \bm \varphi _{21}}{\partial k} + a_{22} - a_{11}\right) - \Im \left[ v^0_{21} \widetilde{\bm v}_{12}^\alpha \right] \frac{\partial \ln | \left\langle 1| {\bm u} \right |2\rangle |}{\partial k} + \Re \left[v^0_{12} \left( a_{22'}{\bm u}_{2'1} - {\bm u}_{21'}a_{1'1}\right)  \right] \right\rbrace +\\ &+ \left\langle 1\leftrightarrow 1' \right\rangle + \left\langle 2 \leftrightarrow 2' \right\rangle + \left\langle 1,2 \leftrightarrow 1',2' \right\rangle,
	\end{split}
	\end{equation}
where $(\bm \varphi)_i$ is the phase of the corresponding $i$-th vector component of the  staggered velocity ${\bm u}$. 
One can perform an SVD transformation of the velocity  matrix in Eq. \eqref{res} in order to diagonalize the velocity matrix. In this case  one of the unitary matrices act in the space of wavefunctions $(1,1')$ and another in space $(2,2')$.  Then the expression \eqref{res} transforms into:
	\begin{equation}
	\begin{split} \label {curr1}
	\bm j^{(1)}_{\text {orb}}  &=   - 2\frac{\Gamma |A|^2  }{\Delta_\Omega^2 + \Gamma^2}   \Im \left [v^0_{21}\left ( \frac{\partial {\bm u}}{\partial k}   \right )_{12} + v^0_{2'1'}\left ( \frac{\partial {\bm u}}{\partial k}   \right )_{1'2'}\right ], \\
	\bm j^{(2)}_{\text{orb}} &= 
	2 \frac{\left( \epsilon_2 - \epsilon_1 - \Omega \right)  |A|^2  }{\Delta_\Omega^2 + \Gamma^2} \Re\left [v^0_{21}\left ( \frac{\partial {\bm u}}{\partial k}   \right )_{12} + v^0_{2'1'}\left ( \frac{\partial {\bm u}}{\partial k}   \right )_{1'2'}\right ].
	\end{split}
	\end{equation}
This resembles the result from Ref.~\onlinecite{Morimotoe1501524}. In terms of the Berry connection Eq.~\eqref{curr1} can be rewritten as 
	\begin{equation} \label{}
	\bm j^{(1)}_{\text{orb}} =   \frac{ \frac{\Gamma}{2} |A|^2  }{\Delta_\Omega^2 + \frac{\Gamma^2}{4}} \left\lbrace  \Re\left[ v^0_{21}{\bm u}_{12}\right]  \left( \frac{\partial \bm \varphi _{21}}{\partial k} + a_{22} - a_{11}\right) - \Im \left[ v^0_{21} {\bm u}_{12}\right] \frac{\partial \ln | \left\langle 1| {\bm u} \right |2\rangle |}{\partial k}  \right\rbrace  + \left\langle 1,2 \leftrightarrow 1',2' \right\rangle
	\end{equation}

Neglecting spin-orbit coupling (i.e. $g_{1,2,3} = 0$), we find that the shift current is given by the last expression in the curly brackets: $\bm j^{(1)}_{\text{orb}} \propto    \Im \left[ v^0_{21} {\bm u}_{12}\right] \frac{\partial \ln | \left\langle 1| {\bm u} \right |2\rangle |}{\partial k}   + \left\langle 1,2 \leftrightarrow 1',2' \right\rangle$. To relate the matrix elements in the last expression with more familiar operators $\tau_x$  and $ \bm v$, one can further expand it to obtain $\bm j^{(1)}_{\text{orb}} \propto    \left ( v_{11}^0 + v_{22}^0\right ) v_{21} \tau^x_{12} \frac{\partial \ln | \bm u _{12}|}{\partial k}   + \left\langle 1,2 \leftrightarrow 1',2' \right\rangle$.

\subsection{Arbitrary direction of polarization}
	
In the case when the light is polarized in the direction that we denote as $\alpha$ and  the orbital  current is measured in direction $\beta$ the expression yields:
	\begin{equation} 
	\begin{split} 
	j^{(1)}_{\text{orb}} =
	-\frac{ \frac{\Gamma}{2} |A|^2  }{\Delta_\Omega^2 + \frac{\Gamma^2}{4}}  	\frac{4}{ \epsilon  }
	\left\lbrace \left( \partial_{k_\beta }\partial_{k_\alpha } d_0\right)  \left[ \left( \partial_{k_\alpha } d_1\right)  d_2 - d_1 \left( \partial_{k_\alpha } d_2\right) \right] - \varepsilon_{lmn}g_l \left(\partial_{k_\alpha } g_m \right) \left( \partial_{k_\beta  }\partial_{k_\alpha } g_n\right)   \right\rbrace 
	\end{split}
	\end{equation}
	
	\subsection{BHZ model  energies and wavefunctions}

The requirements that are imposed on the coefficients in the Hamiltonian due to TR- and I-symmetries are: 
	\begin{equation} \label{d_par}
	\begin{split}
	&d_{0,1} (\bm k) = d_{0,1} (- \bm k) \\
	&d_{2,3,4,5} (\bm k) = -d_{2,3,4,5} (- \bm k) \\
	\end{split}
	\end{equation}
	
The energies and wavefunctions of the Hamiltonian  \eqref{genH}:
	\begin{equation} 
	\epsilon_{1,2} (\bm k)= d_0(\bm k) \pm \sqrt{\sum_{i=1}^5 |d_i (\bm k)|^2 } = d_0(\bm k) \pm \epsilon(\bm k)
	\end{equation}

	\begin{equation} \label{wfgen}
	\begin{split}
	&\psi_1 = \frac{1}{\sqrt{2 \epsilon\left( \epsilon + d_1\right)  }}\begin{pmatrix}
	d_5 - i d_2\\
	d_3 + i d_4\\
	-(\epsilon + d_1)\\
	0
	\end{pmatrix}, \quad 
	\psi_{1'}  = \hat T \hat I \psi_{1} = \frac{1}{\sqrt{2 \epsilon\left( \epsilon + d_1\right)  }}\begin{pmatrix}
	d_3 - i d_4\\
	-(d_5 + i d_2)\\
	0\\
	-(\epsilon + d_1)
	\end{pmatrix},
	\\
	&\psi_{2} = \frac{1}{\sqrt{2 \epsilon\left( \epsilon -d_1\right)  }}\begin{pmatrix}
	d_5- i d_2\\
	d_3 + i d_4\\
	\epsilon - d_1\\
	0
	\end{pmatrix}, \quad 
	\psi_{2'}  = \hat T \hat I \psi_{2} = \frac{1}{\sqrt{2 \epsilon\left( \epsilon - d_1 \right)  }}\begin{pmatrix}
	d_3 - i d_4\\
	-(d_5 + i d_2)\\
	0\\
	\epsilon - d_1
	\end{pmatrix}.\ 
	\end{split}
	\end{equation}
At the topological phase transition, where upon smoothly varying some of the constants in the Hamiltonian the parameter $d_1$ crosses zero and other stay approximately unchanged, the property \eqref{wfs} is easy to see from the explicit form of the wavefunctions.

\subsection{Joint density of states}
	
		\begin{figure*}[t]
			\centering
			\includegraphics[width=0.5\textwidth]{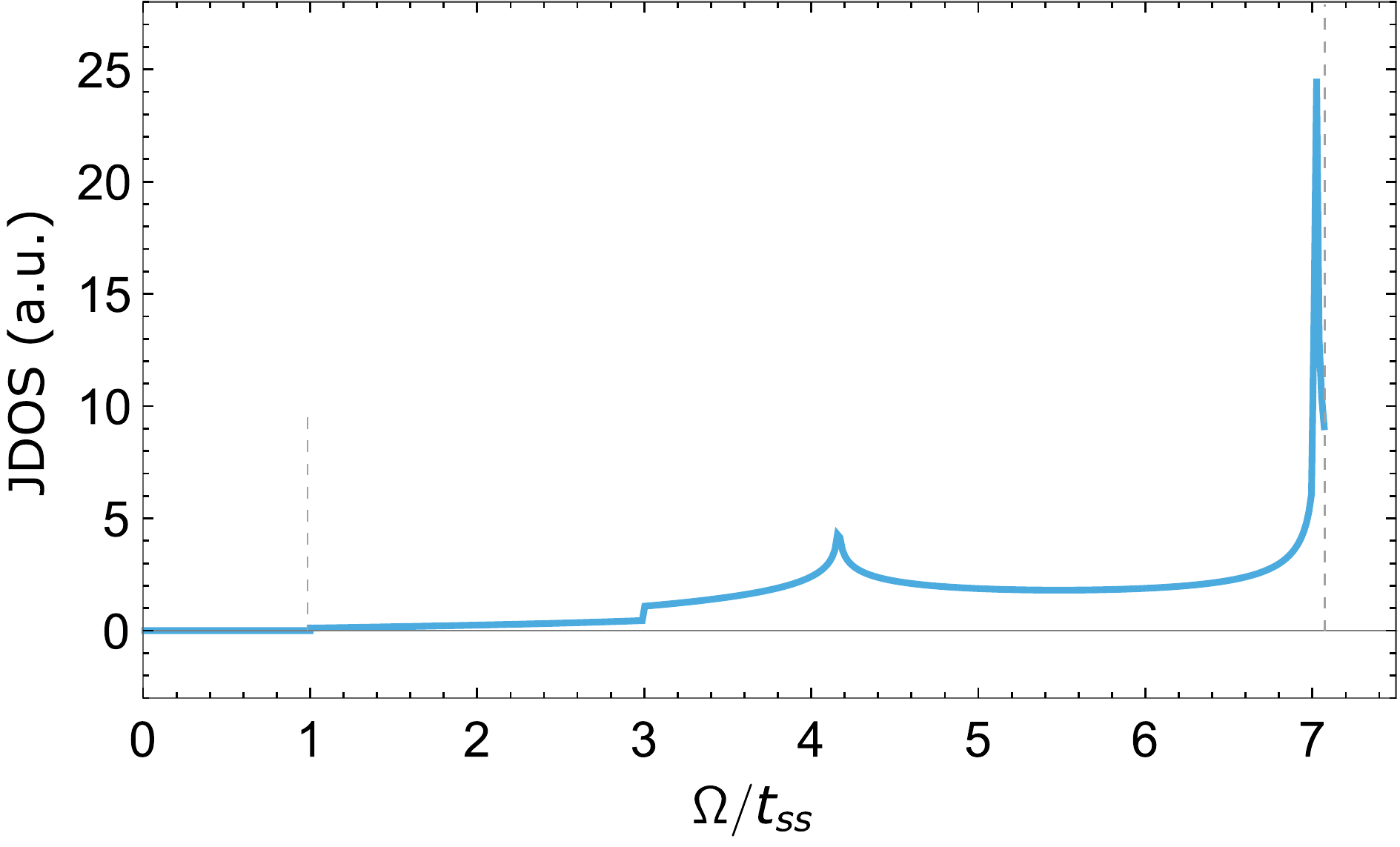}
		\caption{The dependence of the joint density of states  on the light frequency for the BHZ model. $\epsilon_s/t_{ss} = 3$.  The  dashed gray lines correspond to the minimum and maximum frequencies. The other parameters are  $\epsilon_p = t_{pp} = 0$, $t_{sp}/t_{ss} = 1$.
		}
	\end{figure*}
	
 Here we show the joint density of states for parameters similar to those in Fig. \ref{fig2}a. The JDOS reproduces some of the features that are seen in Fig. \ref{fig2}a, namely a jump at $\epsilon_s - \epsilon_p$, and two van Hove singularities at $\Omega/t_{ss} \approx 4.2$ and $\Omega/t_{ss} \approx 7$.

\end{document}